\documentclass[t11pt,preprint,preprintnumbers]{revtex4}
\usepackage{amsmath}
\usepackage{amssymb}
\usepackage{amsfonts}
\usepackage{xcolor}
\usepackage{xfrac}

\newcommand{\bra}[1]{\left\langle #1 \right|}
\newcommand{\ket}[1]{\left|#1\right\rangle}

\newcommand{\overbar}[1]{\mkern 1.5mu\overline{\mkern-1.5mu#1\mkern-1.5mu}\mkern 1.5mu}

\begin{document}
\title{A Quantum Approximate Optimization Algorithm Applied to a Bounded Occurrence Constraint Problem}
\author{Edward Farhi}
\affiliation{Center for Theoretical Physics\\ Massachusetts Institute of Technology\\ Cambridge, MA 02139}
\author{Jeffrey Goldstone}
\affiliation{Center for Theoretical Physics\\ Massachusetts Institute of Technology\\ Cambridge, MA 02139}
\author{Sam Gutmann}

\preprint{MIT-CTP/4628}

\begin{abstract}
We apply our recent Quantum Approximate Optimization Algorithm \cite{farhi-2014} to the combinatorial problem of bounded occurrence Max E3LIN2. The input is a set of linear equations each of which contains exactly three boolean variables and each equation says that the sum of the variables mod 2 is 0 or is 1. Every variable is in no more than $D$ equations.  A random string will satisfy 1/2 of the equations.  We show that the level one QAOA will efficiently produce a string that satisfies $\left(\frac{1}{2} + \frac{1}{101 D^{1/2}\, l n\, D}\right)$ times the number of equations.  A recent classical algorithm \cite{Boaz-2015, hastad-2015} achieved  $\left(\frac{1}{2} + \frac{constant}{D^{1/2}}\right)$.  We also show that in the typical case the quantum computer will output a string that satisfies $\left(\frac{1}{2}+ \frac{1}{2\sqrt{3e}\, D^{1/2}}\right)$ times the number of equations.  
\end{abstract}

\maketitle

\section{Introduction}
We introduced \cite{farhi-2014} a Quantum Approximate Optimization Algorithm, QAOA, which can be used to find  approximate solutions for  combinatorial optimization problems. The algorithm depends on an integer parameter $p\geq1$ and the approximation improves as $p$ increases. Here we only use the $p=1$ algorithm which we now restate.  The input is an $n$ bit instance of a combinatorial problem specified by an objective function $C(z)$ where $z$ is an $n$ bit string and $C(z)$ counts the number of constraints satisfied by the string $z$.  The algorithm works in the $2^n$ dimensional Hilbert space spanned by the computational basis states  $| z \rangle$.  In this basis the objective function $C$ can be viewed as a diagonal operator,
\begin{equation}\label{eq1}
C | z \rangle = C(z) | z \rangle  .
\end{equation}
We also use the operator $B$ which is the sum of the $\sigma_x$ operators,
\begin{equation}\label{eq2}
B = X_1 + X_2 + \ldots + X_n .
\end{equation}
Take as the initial state 
\begin{equation}\label{eq3}
\begin{split}
| s \rangle &= \frac{1}{2^{n/2}} \sum_{z}   | z \rangle \\
&= | + \rangle_1 | + \rangle_2 ... | + \rangle_n
\end{split}
\end{equation}
\noindent which we note is an eigenstate of each of the $X_a$.  Given parameters $\gamma$ and $\beta$, we define the state
\begin{equation}\label{eq4}
|\gamma, \beta \rangle = e^{-i \beta B}e^{-i \gamma C}  | s \rangle.
\end{equation}
The quantum computer is used to produce the state $|\gamma, \beta \rangle$  which is then measured in the computational basis to produce a string $z$.  The unitary operator $e^{-i \beta B}$ is a product of $n$ one qubit operators.  The operator $e^{-i \gamma C}$ can be written as a product of commuting unitaries each of which comes from a constraint in $C$ and has the same locality as the corresponding constraint.  So the number of gates required to produce $|\gamma, \beta\rangle$  is no more than $n$ plus the number of constraints. 

Note that with $\gamma$ and $\beta$ both equal to zero we get a random string and the algorithm is equivalent to this classical algorithm: Pick a string $z$ at random and evaluate $C(z)$.  For some non-zero $\gamma$ and $\beta$ we can do better. In \cite{farhi-2014}  we showed how to efficiently choose $\gamma$ and $\beta$ optimally. For the problem MaxCut we were able to show that on any 3-regular graph the quantum algorithm improves the approximation ratio from $1/2$ (guessing) to $.6924$.  We now look at another problem where we improve on guessing.

Consider the combinatorial problem Max E3LIN2 over $n$ bits.  The E3 means that each clause contains exactly 3 variables.  The LIN2 means that each constraint is a linear equation mod 2 so for say bits $x_1$, $x_2$ and $x_3$ the constraint is either $x_1+x_2+x_3=0$ or $x_1+x_2+x_3=1$. It is possible using Gaussian elimination to determine if the set of linear equations has a solution.  The computational task is to maximize the number of satisfied equations in the case when the equations do not have a solution.  Guessing a random string will satisfy $1/2$ of the equations.  For general instances there is no efficient $\left(\frac{1}{2} + \epsilon\right)$ classical approximation algorithm unless P=NP \cite{hastad-2001}.  

We now make the restriction that every bit is in no more than $D+1$ equations. (The +1 is for later convenience.)  In 2000, H\aa stad \cite{hastad-2000} gave a classical algorithm that achieves an approximation ratio of $\left(\frac{1}{2} + \frac{constant}{D}\right)$.  In 2001, Trevisan \cite{trevisan-2001} showed that the existence of an efficient classical algorithm that achieves an approximation ratio of $\left(\frac{1}{2} + \frac{constant}{D^{1/2}}\right)$, for a sufficiently large constant, would imply that P=NP.  In 2014, the first version of this paper showed that the $p=1$ QAOA applied to an instance of E3LIN2 with $m$ equations produces a string that satisfies at least $\left(\frac{1}{2} + \frac{1}{22 D^{3/4}}\right)m$ equations.  This existential result implies that the approximation ratio is at least $\left(\frac{1}{2} + \frac{1}{22 D^{3/4}}\right)$.  In 2015 a  classical algorithm was discovered \cite{Boaz-2015, hastad-2015} that efficiently finds a string satisfying $\left(\frac{1}{2} + \frac{constant}{ D^{1/2}}\right)m$ equations.  Here we improve the analysis of our earlier version to show that the QAOA produces a string that satisfies $\left(\frac{1}{2} + \frac{1}{101 D^{1/2} \, ln D}\right)m$ equations. We also show that in the typical case the quantum algorithm outputs a string that satisfies $\left(\frac{1}{2}+ \frac{1}{2\sqrt{3e}\, D^{1/2}}\right) m$ equations. By ``typical" we mean this: For any collection of triples  specify an instance of E3LIN2 by picking the $m$ equations associated with the triples to sum to 0 or 1 with 50\%  probability each.
%By typical we mean given any collection of triples that specifies an instance of E3LIN2, we pick the $m$ equations that are associated with the triples to $\sum$ to 0 or 1 each with 50\% probability.

\section{The General Case}

For E3LIN2 we can write the objective operator for any three bits $a$,$b$,$c$ as
\begin{equation}\label{eq5}
\tfrac{1}{2}\left(1 \pm  Z_a  Z_b Z_c\right)
\end{equation}
where the $Z$ operators are $\sigma_z$'s  and the $\pm$ in expression \eqref{eq5} corresponds to the two possible choices for the equation associated with bits $a$,$b$ and $c$.  Dropping the additive constant $1/2$ we write the objective operator as
\begin{equation}\label{eq6}
C = \tfrac{1}{2}\ \sum_{a < b <c}  d_{abc}  \; Z_a Z_b Z_c
\end{equation}
\noindent where $d_{abc}$ is 0 if there is no equation involving $a$, $b$ and $c$ and $d_{abc}$ is $+1$ or $-1$ if there is an equation.  We will evaluate  
\begin{equation}\label{eq7}
\langle - \gamma , \beta | C | - \gamma , \beta \rangle 
\end{equation}
\noindent for certain values of $\gamma$ and $\beta$. Using \eqref{eq6} one can see that \eqref{eq7} is an odd function of $\gamma$. We use $-\gamma$ in \eqref{eq7} to minimize the number of minus signs appearing in the equations below. The expected number of satisfied equations is $m/2$  plus \eqref{eq7}.  For $\gamma$ and $\beta$ equal to 0 the expression \eqref{eq7} is 0 corresponding to just guessing a random string.  To do better we need good choices for $\gamma$ and $\beta$. We pick $\beta = \pi/4$ because it simplifies the analysis. We will show that for some $\gamma$ in the range $[-\frac{1}{10 D^{1/2}}, \frac{1}{10 D^{1/2}}]$ we get the $\left(\frac{1}{2} + \frac{1}{101 D^{1/2} \, ln D}\right)$ result. The optimal $\gamma$ in this range can be found by an efficient search.   

Consider one term in the quantum expectation \eqref{eq7} that comes from the clause involving say bits 1, 2 and 3.  This is
\begin{equation}\label{eq8}
\tfrac{1}{2}\ d_{123}  \langle s | e^{- i \gamma C} e^{i \beta B} Z_1 Z_2 Z_3 e^{-i \beta B}e^{i \gamma C} | s \rangle .
\end{equation}
\noindent Now in the $B$ operator all terms except the $X_1 + X_2 + X_3$ commute through the three $Z$'s.  We pick $\beta=\pi/4$ and get 
\begin{equation}\label{eq9}
\tfrac{1}{2}\ d_{123} \langle s | e^{- i \gamma C} Y_1 Y_2 Y_3 e^{i \gamma C} | s \rangle,
\end{equation}
\noindent where the $Y$'s are $\sigma_y$'s.  We separate out the clause involving bits 1,2 and 3 in $C$ and write 
\begin{equation}\label{eq10}
C = \overbar{C} + \tfrac{1}{2}\ d_{123}\, Z_1 Z_2 Z_3 .
\end{equation}
\noindent Conjugating the $Y_1 Y_2 Y_3$ with the contribution from clause 123 we get
\begin{equation}\label{eq11}
\tfrac{1}{2}\ d_{123} \langle s | e^{-i \gamma \overbar{C}} \bigl( \cos(\gamma d_{123})Y_1 Y_2 Y_3 + \sin(\gamma d_{123})X_1 X_2 X_3 \bigr) e^{i \gamma \overbar{C} }| s \rangle .
\end{equation}
\noindent We will first evaluate %the term with three $X$ operators whose coefficient is always $-\frac{1}{2}\sin(\gamma)$,
\begin{equation}\label{eq12}
\langle s | e^{-i \gamma \overbar{C}} X_1 X_2 X_3 e^{i \gamma \overbar{C}} | s \rangle .
\end{equation}
\noindent Insert two complete sets for qubits 1,2 and 3 to get
\begin{equation}
\sum_{z_1,z_2,z_3} \sum_{z_1',z_2',z_3'} \langle s | e^{- i \gamma \overbar{C}} |z_1,z_2,z_3\rangle\langle z_1,z_2,z_3| X_1 X_2 X_3|z_1'z_2',z_3'\rangle \langle z_1',z_2',z_3'|e^{i \gamma \overbar{C}} | s \rangle.
\end{equation}
\noindent Now the $X$ operators are off diagonal so we get for \eqref{eq12}
\begin{equation}\label{eq13}
\sum_{z_1,z_2,z_3} \langle s | e^{-i \gamma \overbar{C}} |z_1,z_2,z_3\rangle \langle -z_1,-z_2,-z_3 | e^{i \gamma \overbar{C}}| s \rangle .
\end{equation}
\indent We now need to look more carefully at $\overbar{C}$ which includes all the clauses involving bits 1,2 and 3 except the 1,2,3 constraint. We assume that the instance has bounded degree, that is, that each bit is in no more than $D+1$ clauses.  So aside from the central clause each of bits 1,2 and 3 can be in at most $D$ clauses so there are at most $3D$ clauses in $\overbar{C}$.  Each of these clauses involves at most two bits besides 1, 2 and 3 so there could be as many as $6D$ bits other than bits 1,2 and 3 involved in the starting expression \eqref{eq8} as well as in \eqref{eq13} but there may be fewer.  Let us write $\overbar{C}$ as
\begin{equation}
\overbar{C}=Z_1C_1 + Z_2C_2 + Z_3C_3 + Z_1Z_2C_{12} + Z_1 Z_3 C_{13} + Z_2 Z_3 C_{23}
\label{eq14}
\end{equation}
\noindent where $C_1$ is a sum of terms of the form  $\frac{1}{2}\, d_{1ab}\, Z_a  Z_b$ where $a$ and $b$ are pairs of bits other than 2 and 3 that come from clauses with bit 1.  Similarly for $C_2$ and $C_3$.   $C_{12}$ is a sum of terms of the form $\frac{1}{2}\, d_{12a}\, Z_a$ where bit $a$ is not  3 and comes from being in a clause with bits 1 and 2. Similarly for $C_{13}$ and $C_{23}$.    Note in expression \eqref{eq13} there is a $-\overbar{C}$ on the left where bits 1, 2 and 3 take the values $z_1$, $z_2$ and $z_3$ whereas on the right we have $\overbar{C}$ with the bits taking the values $-z_1$,$-z_2$ and $-z_3$.  So expression \eqref{eq13} can be written as
\begin{equation}
\tfrac{1}{8}\sum_{z_1,z_2,z_3}   \langle \overbar{s} | e^{-2 i \gamma \left(z_1 C_1+ z_2 C_2+z_3 C_3\right)} | \overbar{s} \rangle
\label{eq15}
\end{equation}
\noindent where
\begin{equation}\label{eq16}
|\overbar{s}\rangle = \prod_{a\in\mathcal{Q}}  | + \rangle_a 
\end{equation}
and $a$ is in the set $\mathcal{Q}$ consisting of qubits that appear in $C_1$, $C_2$ and $C_3$.  Note that the number of elements in $\mathcal{Q}$, $Q$, can be as large as $6D$.  We can do the sum in \eqref{eq15} explicitly and we get
\begin{multline}\label{eq17}
\frac{1}{4} \langle \overbar{s} | \bigl[\cos(2 \gamma (C_1+C_2+C_3)) + \cos(2\gamma (C_1-C_2-C_3))\\  
+\cos(2 \gamma (-C_1+ C_2 - C_3)) +\cos(2 \gamma (- C_1 - C_2 + C_3))\bigr] | \overbar{s} \rangle .
\end{multline}
\noindent Now we write 
\begin{equation}\label{eq18}
C_i = \tfrac{1}{2}\sum_{a<b}\, d_{iab}\; Z_a Z_b \; \; , \; \;i = 1,2,3.
\end{equation}
\noindent    We can write the quantum expectation in \eqref{eq17} as %and $J^{(1)}$, $J^{(2)}$ and $J^{(3)}$ have entries of $1$ or $-1$ if the pair $a,b$ comes from a $+$ or $-$ clause and the entry is 0 if the pair $a, b$ is not associated with a clause.
\begin{multline}\label{eq19}
\frac{1}{4\cdot2^Q} \sum_{\left\{z_a\right\}\, a \epsilon \mathcal{Q}}\bigl[\cos(\gamma (c_1(z)+c_2(z)+c_3(z)))+\cos(\gamma (c_1(z)-c_2(z)-c_3(z)))\\ 
+\cos(\gamma (-c_1(z) + c_2(z) - c_3(z)))+\cos(\gamma (-c_1(z)- c_2(z)+ c_3(z)))\bigr]
\end{multline}
\noindent where      
\begin{equation}\label{eq20}
c_i(z)= \sum_{a < b}\, d_{iab}\, z_a  z_b \; \; , \; \; i=1,2,3. 
\end{equation}
\noindent It is convenient for us to view $c_1$, $c_2$ and $c_3$ as random variables coming from an underlying distribution of $Q$ binary variables so we write  \eqref{eq12} as
\begin{multline}\label{eq22}
\tfrac{1}{4}\, \mathbb{E}_z\, [ \cos(\gamma(c_1+c_2+c_3))+\cos(\gamma (c_1-c_2-c_3))\\
+\cos(\gamma (-c_1+ c_2 - c_3))+\cos(\gamma (-c_1 - c_2 + c_3))]  
\end{multline}
where $\mathbb{E}_z$ is a classical expectation over binary variables $z$. The full contribution to \eqref{eq11} from the $X_1 X_2 X_3$ terms is
\begin{equation}\label{eq23}
\tfrac{1}{8}\, d_{123} \sin (\gamma d_{123}) \mathbb{E}_z [\cos \left(\gamma (c_1 + c_2+c_3) \right) + \ldots ] \ .
\end{equation}
To evaluate the $Y_1 Y_2 Y_3$ term in \eqref{eq11} we follow the same steps that led to \eqref{eq23} and get 
\begin{equation}\label{eq24}   
\tfrac{1}{8}\,  d_{123} \cos(\gamma d_{123}) \mathbb{E}_z [\sin (\gamma (c_1+c_2+c_3) + \dots ] \, .
\end{equation}
 Adding \eqref{eq23}  and \eqref{eq24} we see that \eqref{eq11}, and therefore \eqref{eq8}, with $\beta = \pi/4$, is equal to 
 \begin{alignat}{2}\label{eq25}
& \tfrac{1}{8}\, d_{123}\, \mathbb{E}_z\, [\sin (\gamma (d_{123} + c_1 + c_2 + c_3)) 
  + \sin (\gamma (d_{123} + c_1 - c_2 - c_3)) \nonumber\\
 &+ \sin (\gamma (d_{123} - c_1 + c_2 - c_3))
  + \sin (\gamma (d_{123} - c_1 - c_2 + c_3))] .
 \end{alignat}

We now Taylor expand \eqref{eq25} singling  out the linear term to get
\begin{equation}\label{eq26}
\tfrac{1}{2}\; d^2_{123} \gamma + P^k_{123}  (\gamma) + \mathcal{R}^k_{123} (\gamma)
\end{equation}
where
\begin{equation}\label{eq27}
P^k_{123} (\gamma) = \tfrac{1}{8}\; d_{123} \sum\limits^k_{j = 3, 5 \ldots}\ \frac{\gamma^j (-1)^{\frac{j-1}{2}}}{j !}\ \mathbb{E}_z\, [\,(d_{123} + c_1+ c_2 + c_3)^j + \ldots ]
\end{equation}
and
\begin{equation}\label{eq28}
\left| \mathcal{R}^k_{123} (\gamma) \right| \leqslant \frac{1}{8}\ \frac{|\gamma|^{k+2}}{(k+2)!}\; \mathbb{E}_z\, [\, \left|d_{123} + c_1 + c_2+ c_3 \right|^{k+2} + \ldots] .
\end{equation}
Note  that the coefficient of the linear term in \eqref{eq26} is $\frac{1}{2} d^2_{123} = \frac{1}{2}$ regardless of the equation type. The polynomial $P^k_{123}$ starts with a cubic term and is of $k^{th}$ order where $k$, to be chosen shortly, will depend on $D$. The values of the coefficients of $\gamma^j$ in $P^k_{123}$ will play no role in lower bounding the performance of the quantum algorithm. We now bound the right hand side of \eqref{eq28}. 

From \eqref{eq20} we see that $c_1, c_2 \ \text{and}\ c_3$ are degree 2 polynomials in the $z_a$ and so are\\ $d_{123} \pm c_1 \pm c_2 \pm c_3$. By Theorem 5 of \cite{Dinur-2007} we have that for any degree 2 polynomial $c$,
\begin{equation}\label{eq29}
\mathbb{E}_z \left[|c|^{k+2}\right] \leqslant (k+1)^{k+2} \left(\mathbb{E}_z [c^2]\right)^{\frac{k+2}{2}} .
\end{equation}
Now
\begin{equation}\label{eq30}
\mathbb{E}_z \left[(d_{123} \pm c_1 \pm c_2 \pm c_3)^2\right] = 1 + \mathbb{E}_z \left[(c_1 \pm c_2 \pm c_3)^2\right]
\end{equation}
since $\mathbb{E}_z [c_i]=0$. Returning to \eqref{eq20} we see that 
\begin{eqnarray}\label{eq31}
\mathbb{E}_z [c^2_i] = & \sum\limits_{a <b} & d_{iab}\; d_{iab}\\[.5ex]
 \leqslant & D\nonumber
\end{eqnarray}
because of our bounded occurrence assumption.  So (using Cauchy-Schwarz),
\begin{equation}\label{eq32}
\mathbb{E}_z \left[(d_{123} \pm c_1 \pm c_2 \pm c_3)^2\right] \leqslant 1 + 9 D .
\end{equation}
Plugging in to \eqref{eq29} gives
\begin{equation}\label{eq33}
\mathbb{E}_z \left[|d_{123} \pm c_1 \pm c_2 \pm c_3|^{k+2}\right] \leqslant (k+1)^{k+2} (1 + 9D)^{\frac{k+2}{2}}
\end{equation}
and we have that 
\begin{alignat}{4}\label{eq34}
\left|\mathcal{R}^k_{123} (\gamma)\right| &\leqslant \frac{1}{2} \frac{(k+1)^{k+2}}{(k+2)!}\; (1 + 9D)^{\frac{k+2}{2}} & |\gamma|^{k+2}\nonumber\\[.5ex]
& \leqslant  \frac{1}{2} \left(e(1 +9D)^{1/2} |\gamma|\right)^{k+2}&\\[.ex]
&\leqslant \left( 9 D^{1/2} |\gamma|\right)^{k+2}&\nonumber
\end{alignat}
where  we used Stirling's formula and that $e < 3$.

Recall that our goal is to evaluate \eqref{eq7} which is a sum of $m$ terms of the form \eqref{eq25} which can be written, as in \eqref{eq26},
\begin{alignat}{2}\label{eq35}
\langle -\gamma, \pi/4| C | -\gamma,  \pi/4\rangle
&= \sum\limits_{a < b < c} \{\tfrac{1}{2} \;  \gamma + P^k_{abc} (\gamma) + \mathcal{R}^k_{abc} (\gamma)\}\\[.5ex]
& = \frac{m}{2} \; \gamma + P^k (\gamma) + \sum\limits_{a< b < c} \mathcal{R}^k_{abc} (\gamma)\nonumber
\end{alignat}
where $P^k (\gamma)$ is the sum of the $m$ polynomials $P^k_{abc} (\gamma)$. By the triangle inequality 
\begin{equation}\label{eq36}
\bigg|\langle -\gamma, \pi/4| C | - \gamma,  \pi/4\rangle \bigg| \geqslant  \bigg| \frac{m}{2} \; \gamma + P^k (\gamma) \bigg| - \sum\limits_{a < b < c} \bigg| \mathcal{R}^k_{abc} (\gamma) \bigg|
\end{equation}
which by \eqref{eq34} is
\begin{equation}\label{eq37}
\geqslant  \left| \frac{m}{2} \; \gamma + P^k (\gamma) \right| - m\; (9 D^{1/2} |\gamma|)^{k + 2} .
\end{equation}
To keep the negative term in \eqref{eq37} small we take
\begin{equation}\label{eq38}
| \gamma | \leqslant  \frac{1}{10 D^{1/2}} \ .
\end{equation}
To lower bound the positive term in \eqref{eq37} we make use of Corollary 2.7 in \cite{Dinur-2007}:\\
With $x_r = \cos (\pi r/k)$,
\begin{equation}\label{eq39}
\max\limits_{r = 0,1\ldots k} \; \;   \Big| x_r + a_2\;  x^2_r + \ldots a_k\;  x^k_r  \Big| \geqslant \frac{1}{k}%max |x| \leqslant | | x + P_2 x^2 + \ldots P_k x^\k \geqslant \frac{1}{k}
\end{equation}
for any real $a_2 \ldots a_k$ with $k$ odd. This implies that (for odd $k$) with
\begin{equation}\label{eq40-TEMP}%%%%%%%%%%%%%%%%%%%%%%%%%%%%%%%%%%%NEW EQUATION
\gamma_r = \frac{1}{10 D^{1/2}}\ \cos (\pi r/k)
\end{equation}
we have
\begin{equation}\label{eq40}
\max\limits_{r = 0,1\ldots k}  \; \; \left| \frac{m}{2}\; \gamma_r + P^k (\gamma_r) \right| \geqslant \frac{m}{20 D^{1/2} k}\ .
\end{equation}
Returning to \eqref{eq37} we have
\begin{eqnarray}\label{eq41}
\max\limits_{r = 0,1\ldots k} \; \left\{ \left| \frac{m}{2}\,  \gamma_r + P^k (\gamma_r) \right| - m (9 D^{1/2} |\gamma_r |)^{k+2} \right\} \geqslant \frac{m}{20 D^{1/2} k} \; - m \left(\frac{9}{10}\right)^{k+2} .
\end{eqnarray}
Now take  $k = 5\, l n D$ to make the positive term dominate for large $D$ and we have that the right hand side of \eqref{eq41} is greater than
\begin{equation}\label{eq42}
\frac{m}{101 D^{1/2}\, l n D}
\end{equation}
for large $D$.

What we have shown is that there is a value of $\gamma$ between $-\frac{1}{10 D^{1/2}}$ and $\frac{1}{10 D^{1/2}}$ for which the absolute value of \eqref{eq7} with $\beta = \pi / 4$ is greater than \eqref{eq42}. And since \eqref{eq7} is an odd function of $\gamma$ there is a value of $r$ that makes \eqref{eq7} positive, so
\begin{equation}\label{eq43}
 \max\limits_{r = 0,1\ldots k}   \; \; \langle \gamma_r, \pi/4 | C | \gamma_r,  \pi/4 \rangle \geqslant \frac{m}{101 D^{1/2}  l n  D}\ .
\end{equation}
So we need to run the quantum computer at no more than $5 \, ln D$ values of $\gamma$ to achieve \eqref{eq42}.
 
We have shown that for any instance of E3LIN2 with $m$ equations there is a string that satisfies at least
 \begin{equation}\label{eq46}
\left( 
\frac{1}{2} + \frac{1}{101 D^{1/2}\, ln\, D}
\right)\, m
\end{equation}
equations. Running the quantum computer repeatedly produces a sample of strings for which the expected number of equations satisfied is at least \eqref{eq46}. A sample of size $m \log m$ will with probability $1 -\frac{1}{m}$ include a string that satisfies at least this many equations.

 \section{The Typical Case}
 In the previous section we showed that for any instance, the performance of the quantum algorithm is  at least \eqref{eq46}. From \cite{Boaz-2015} we know that for every instance there is a string that satisfies at least $\left(\frac{1}{2} + \frac{constant}{D^{1/2}}\right) m$ equations and the authors give an efficient classical algorithm that produces such a string. We do not know if our quantum algorithm achieves this for all instances, but it does achieve it on ``typical" instances as we now discuss.
 
 Every instance of E3LIN2 is specified by a collection of $m$ triples $a, b, c$ and for each triple a choice of 0 or 1 specifying the equation to be
 \begin{equation}\label{eq47}
(x_a + x_b + x_c)\negthickspace\negthickspace\negthickspace \mod 2 = \begin{cases} 0\\ 1 \end{cases} .
\end{equation}
Thus for each fixed collection of triples there are $2^m$ instances. For any  fixed collection of triples we now determine the performance of the algorithm on an instance uniformly selected from the $2^m$ possible instances.  Return to equation \eqref{eq6}.  Each $d_{abc}$ which is not 0 is now independently chosen to be $+1$ or $-1$ with probability $1/2$. With this distribution let us calculate the expected value of \eqref{eq9} where again $\beta = \pi/4$ but $\gamma$ is not yet specified. Recall that \eqref{eq9} is the sum of the two contributions \eqref{eq23} and \eqref{eq24}. Since $c_1, c_2 \ \text{and} \ c_3$ do not involve $d_{123}$, the expected value of \eqref{eq24} over $d_{123}$ is $0$. Using that $d_{123}$ is $+1$ or $-1$ we can write \eqref{eq23} as
\begin{multline}\label{eq48}
\tfrac{1}{8}\, (\sin \gamma)\, \mathbb{E}_z [\cos\, (\gamma(c_1+c_2+c_3))+\cos\, (\gamma (c_1-c_2-c_3))\\
+\cos\, (\gamma (-c_1+ c_2 - c_3))+\cos\, (\gamma (-c_1 - c_2 + c_3))] \\
=  \tfrac{1}{2}\, (\sin \gamma)\,  \mathbb{E}_z [ \cos\, (\gamma c_1) \cos\, (\gamma c_2) \cos\, (\gamma c_3)]\ .
\end{multline}
We now evaluate the expected value $\mathbb{E}_d$ of \eqref{eq48}. First note that
\begin{eqnarray}\label{eq49}
\mathbb{E}_d \left[\cos\, (\gamma c_1)\right] &=& \mathbb{E}_d \left[\cos\, \bigg(\gamma \sum\limits_{3 < a <b}\, d_{1ab}\, z_a z_b \bigg)\right]\nonumber\\[.5ex]
%&=& \frac{1}{2}\ \mathbb{E}_d \left[\exp\, \bigg(i \gamma \sum\limits_{3 < a <b} d_{1ab}\, z_a z_b\bigg) + \exp \bigg(- i \gamma \sum\limits_{3 < a <b} d_{1ab}\, z_a z_b\bigg) \right]\nonumber \\[.5ex]
&=& \frac{1}{2}\ \mathbb{E}_d \left[\prod\limits_{3< a<b} \exp \left(i \gamma\, d_{1ab}\, z_a z_b\right) +  \prod\limits_{3< a<b} \exp \left(- i \gamma\, d_{1ab}\, z_a z_b\right) \right]\nonumber\\[.5ex]
&=& \prod\limits_{3< a<b} \cos\, (\gamma\, z_a z_b)\nonumber\\[.5ex]
&=& \prod\limits_{3< a<b} \cos\, \gamma
\end{eqnarray}
where it is understood that the product is only over values of $a$ and $b$ where $d_{1ab}$ is not $0$. Note that \eqref{eq49} does not depend on the $z$'s. Plugging into \eqref{eq48} we see that (since $c_1$, $c_2$, and $c_3$ involve distinct $d$'s) the expected value of \eqref{eq48} over the $d$'s is
\begin{equation}\label{eq50}
\frac{1}{2} (\sin \gamma) (\cos \gamma)^{(D_1 + D_2 +D_3)}
\end{equation}
where $D_i$ is the number of terms in $c_i$ so $0 \leqslant (D_1 + D_2 +D_3) \leqslant 3D$.

 Summing the contributions from all clauses we have
\begin{equation}\label{eq51}
\frac{m}{2} \sin \gamma \cos^{3D} \gamma \leqslant \mathbb{E}_d \Big[\bra{- \gamma,  \pi/4} C \ket{- \gamma,  \pi/4}\Big] \leqslant \frac{m}{2} \sin \gamma\ .
\end{equation}
We want to choose $\gamma$ to maximize the lower bound. Let
\begin{equation}\label{eq52}
\gamma = \frac{g}{D^{1/2}} \ .
\end{equation}
For large $D$ the lower bound in \eqref{eq51}  is 
\begin{equation}\label{eq53}
\frac{m}{2}\ \frac{g}{D^{1/2}} \ \exp \left(-\frac{3}{2} \, g^2 \right)
\end{equation}
which is maximized when $g = \frac{1}{\sqrt{3}}$ giving a lower bound of
\begin{equation}\label{eq54}
\frac{m}{2\sqrt{3e}\, D^{1/2}} \ .
\end{equation}
So averaging over $d$'s we have that with $\gamma = 1/\sqrt{3}\, D^{1/2}$, the expected number of satisfied equations is  at least
\begin{equation}\label{eq55}
\left(\frac{1}{2} + \frac{1}{2\sqrt{3e}\, D^{1/2}} \right) \, m.
\end{equation}

To see that the average over the $d$'s is the typical performance for the $2^m$ instances associated with a given collection of triples, we need to look at the variance. At $\beta=\pi/4$, our target \eqref{eq7} is the sum of $m$  terms
\begin{equation}\label{eq55-TEMP}%%%%%%%%%%%%%%%%%%%%%%%%%%%%%%%%%%%%NEW EQUATION
\bra{-\gamma,  \pi/4} C \ket{-\gamma,  \pi/4} = \sum\limits_{a < b <c} \, \tfrac{1}{2} \ d_{abc} \bra{-\gamma,  \pi/4} Z_a Z_b Z_c \ket{-\gamma,  \pi/4} 
\end{equation}
whose expected value over the $d$'s we have just evaluated. For the variance we need the square which consists of a sum of terms of the form
\begin{equation}\label{eq56-TEMP}%%%%%%%%%%%%%%%%%%%%%%%%%%%%%%%%%%%%NEW EQUATION
 d_{abc} \bra{-\gamma,  \pi/4} Z_a Z_b Z_c \ket{-\gamma,  \pi/4} \cdot d_{a^\prime b^\prime c^\prime} \bra{-\gamma,  \pi/4} Z_{a^\prime} Z_{b^\prime} Z_{c^\prime} \ket{-\gamma,  \pi/4}
\end{equation}
where each factor is of the form \eqref{eq25} with 123 replaced by $abc$ or $a^\prime b^\prime c^\prime$. Note that in \eqref{eq25} $c_1$, only involves $d$'s which have one subscript equal to 1 and similarly for $c_2$ and $c_3$. Now if all of the $d$'s that appear in the expression \eqref{eq25} with $abc$ are distinct from all of the $d$'s that appear in \eqref{eq25} with $a^\prime b^\prime c^\prime$ then
\begin{multline}\label{eq57-TEMP}%%%%%%%%%%%%%%%%%%%%%%%%%%%%%%%%%%%%NEW EQUATION
\mathbb{E}_d \Big[d_{abc} \bra{-\gamma,  \pi/4} Z_a Z_b Z_c \ket{-\gamma,  \pi/4}\cdot  d_{a^\prime b^\prime c^\prime} \bra{-\gamma,  \pi/4} Z_{a^\prime} Z_{b^\prime} Z_{c^\prime} \ket{-\gamma,  \pi/4} \Big]\\
= \mathbb{E}_d \Big[d_{abc} \bra{-\gamma,  \pi/4} Z_a Z_b Z_c \ket{-\gamma,  \pi/4}\Big] \cdot \mathbb{E}_d \Big[d_{a^\prime b^\prime c^\prime} \bra{-\gamma,  \pi/4} Z_{a^\prime} Z_{b^\prime} Z_{c^\prime} \ket{-\gamma,  \pi/4} \Big]\, .
\end{multline}
This means that the only contributions to the variance can come from terms where at least one of the $d$'s in the $abc$ term also appears in the $a^\prime b^\prime c^\prime$ term. We now count the maximum number of times that this can occur.

Fix $abc$. The full $abc$ factor in \eqref{eq56-TEMP} involves bits $a, b$ and $c$  and at most $6 D$ other bits for a total of at most ($6D + 3$) bits. Each of these bits is in at most ($D+1$) clauses so $abc$ is ``linked" to at most $(6D+3)(D+1)$ clauses. Unless $a^\prime b^\prime c^\prime$ is one of these ``linked" clauses then \eqref{eq57-TEMP} is guaranteed to be satisfied. Therefore there are at most $m(6 D+3) (D+1)$ possible contributions to the variance. Returning to \eqref{eq57-TEMP} we see that each term in the sum is between $-1/2$ and $+1/2$ so the variance is at most $\frac{1}{4} m (6D+3) (D+1)$.

The mean (over $d$'s) of the number of satisfied equations produced by the quantum algorithm is of order $m$ and we just showed that the standard deviation is of order $\sqrt{m}$. (Recall that we keep $D$ fixed and let $m$ be large.) Thus for large $m$, with high probability, on a ``typical" instance, the quantum algorithm outputs a string satisfying at least \eqref{eq55} minus a term of order $\sqrt{m}$.
 
\section{Conclusions}
We applied the Quantum Approximate Optimization Algorithm at level $p=1$ with predetermined values of $\gamma$ and $\beta$ to the problem of Max E3LIN2 with each bit in no more than $D+1$ equations.  We have shown that the quantum computer will output a string that satisfies 
\begin{equation}
 \left(\frac{1}{2} + \frac{1}{101 D^{1/2}\, ln \, D}\right) m
\end{equation}
equations.
 
An instance of  E3LIN2 is specified by a collection of $m$ triples and a choice of 0 or 1 for the equation associated with each triple. If 51\% of these choices are 1 then the all 1's string  will satisfy 51\% of the equations independent of $D$. Therefore the hard cases must have nearly equal numbers of 0's and 1's. We showed that for {\it any} collection of triples if the 0's and 1's are chosen at random (50/50) then the algorithm satisfies 
\begin{equation}\label{eq56}
\left(\frac{1}{2}+ \frac{1}{2\sqrt{3e}\, D^{1/2}}\right) m
\end{equation}
equations with high probability. It would be interesting to find an instance where the algorithm does not achieve \eqref{eq56}.

        The performance of the quantum algorithm can be improved.  Here are a number of ways:\vspace{-1.25ex}
\begin{itemize}
\item We picked $\beta=\pi/4$ for ease of analysis and  a small set of values of $\gamma$ which sufficed for our results.  Instead at $p=1$ one could search for          the optimal values of $\beta$ and $\gamma$ for each input instance. This could be done by classical preprocessing or by hunting for the best $\beta$ and $\gamma$ by making calls to the quantum computer.
\item At $p=1$ we could expand the parameter space. For example we could have a different angle $\gamma$ for each clause. This could only                   improve performance.
\item Go to higher $p$.  Perhaps at a higher value of $p$, the dependence of the approximation ratio on $D$ will be better than $\frac{constant}{D^{1/2}\, ln \, D}$. 
\end{itemize}
\section{Acknowledgements}

We thank Oded Regev,  Madhu Sudan,  Luca Trevisan and Salil Vadhan for helpful correspondence.  SG thanks Luca Trevisan for a stimulating coffee. We thank Boaz Barak and Ankur Morita for discussion.  We are indebted to Elizabeth Crosson and Charles Suggs for help in preparing the manuscript.  This work was supported by the US Army Research Laboratory's Army Research Office through grant number W911NF-12-1-0486, and the National Science Foundation through grant number CCF-121-8176.

\end{document}